\documentclass[prd, twocolumn, nofootinbib, floatfix]{revtex4}

\usepackage{epsfig} 
\usepackage{amsmath}
\usepackage{amsbsy}

\newcommand{\beq}{\begin{equation}}
\newcommand{\eeq}{\end{equation}}
\newcommand{\beqa}{\begin{eqnarray}}
\newcommand{\eeqa}{\end{eqnarray}}
\newcommand{\om}{\Omega_m}

\newcommand{\wm}{W_-}
\newcommand{\wpcap}{W_+}
\newcommand{\caln}{{\mathcal N}}
\newcommand{\dc}{\Delta\chi} 
\newcommand{\dcc}{\Delta\chi^2} 
 
\newcommand{\ls}{\mathrel{\raise0.27ex\hbox{$<$}\kern-0.70em \lower0.71ex\hbox{{
$\scriptstyle \sim$}}}}

\begin{document} 

\title{Generating and Analyzing Constrained Dark Energy Equations of State 
and Systematics Functions}
\author{Johan Samsing$^{1}$ \& Eric V.\ Linder$^{2,3}$}
\affiliation{$^1$Dark Cosmology Centre, Niels Bohr Institute, 
University of Copenhagen, Copenhagen, Denmark \\ 
$^2$Berkeley Lab \& University of California, Berkeley, CA 94720, USA \\ 
$^3$Institute for the Early Universe, Ewha Womans University, Seoul, Korea} 
\date{\today}

\begin{abstract} 
Some functions entering cosmological analysis, such as the dark 
energy equation of state or systematic uncertainties, are 
unknown functions of redshift.  To include them without assuming a 
particular form we derive an efficient method 
for generating realizations of all possible functions subject to certain 
bounds or physical conditions, e.g.\ $w\in[-1,+1]$ as for quintessence.  
The method is optimal in the sense that it is both pure and complete in 
filling the allowed space of principal components.  
The technique is applied to propagation of systematic uncertainties in 
supernova population drift and dust corrections and calibration through 
to cosmology parameter estimation and bias in the magnitude-redshift 
Hubble diagram.  
We identify specific ranges of redshift and wavelength bands 
where the greatest improvements in supernova systematics due to population 
evolution and dust correction can be achieved. 
\end{abstract} 

\maketitle

\section{Introduction \label{sec:intro}}

The nature of the dark energy accelerating the cosmic expansion is a 
major mystery of modern physics.  The effectively negative pressure 
giving rise to acceleration can be parametrized through the equation of 
state, or pressure to energy density, ratio of the dark energy.  
Observational quantities such as the distance-redshift relation, 
Hubble expansion rate, or matter density perturbation growth (assuming 
general relativity) can then be derived in terms of the equation of 
state (EOS).  However, little guidance exists from theory for the form 
of the EOS. 

One of the standard approaches is to adopt a well-tested, nearly unbiased 
functional form for the EOS.  For example, the EOS as a function of scale 
factor, $w(a)=w_0+w_a(1-a)$, where $w_0$, $w_a$ are parameters to be fit, 
has been shown to be accurate at the $0.1\%$ level in the observable 
distance for a wide array of 
dark energy models \cite{calib}.  However, one may prefer to keep the 
EOS as free as possible.  Values in bins of redshift, or 
some form of eigenmodes or principal components, do not impose assumptions 
on the form of $w(a)$ (see, e.g., \cite{hutstar,coohut,tobin,mhh,kitching,mort09}). 

Physics does bound the possible behaviors, though, not allowing full 
freedom in the bin values or principal component coefficients.  One 
example involves a minimally coupled, canonical scalar field, where 
for all redshifts the condition must hold that $w\in[-1,+1]$ for 
positive energy density.  
Principal components can be applied to many situations, such as the 
cosmic reionization fraction history or the fraction of a source 
population in a particular subclass, where values can only lie in $[0,1]$. 
We generically call such unknown, redshift dependent quantities 
``state functions''. 
Some approaches to such situations of ``freedom under constraint'' 
exist in the cosmology literature, e.g.\ \cite{mort1,tobin,mhh,lin0812,amara}, 
but here we concentrate on full and computationally efficient solutions. 

Going further into the motivation, we consider three reasons for imposing 
bounds: physicality, efficiency, and prior information.  Some physical 
bounds are absolute, such as an ionization fraction ranging between 0 
and 1, while others are more relative, such as the dark energy equation 
of state ranging between $-1$ and $+1$ for a canonical, minimally coupled 
scalar field.  In fact, there is a certain amount of framework dependence 
in any analysis -- the matter density cannot be less than zero, but the 
effective matter density can appear less than zero when a universe with a 
cosmological constant is interpreted in terms of a pure matter universe: 
this is precisely how the acceleration of the universe was discovered. 
So while physical bounds are generally valid, results pushing up against 
the bounds should sound a note of caution; one might then loosen the 
bounds to check for consistent results.  But starting with overly loose 
or unmotivated bounds 
has the price of computational 
inefficiency; in the vast majority of cases one would not scan over a 
space where the ionization fraction ranged from $-5$ to $+5$, say.  
Finally, the bounds may arise from prior information such as having 
measured a calibration offset to be less than some value (as we apply 
in Sec.~\ref{sec:dust}).  There is little point in examining the effect 
of larger variations than allowed by this prior information.  These 
rationales for bounds on the state function then translate directly 
into the principal component space.  We emphasize that only the 
amplitude, not the freedom in the functional form, is being limited. 

When selecting physically valid principal component contributions the 
two main issues are those of purity -- every set of values gives a 
valid state function -- and completeness -- every possible valid 
state function is represented in the selection.  In Sec.~\ref{sec:method} 
we discuss possible methods for generating principal component 
realizations of the EOS (or any other) function and assess their 
purity and completeness, especially when only a subset of modes is 
retained.  We present a solution for the optimal -- pure and complete -- 
prescription in Sec.~\ref{sec:hypercube}, along with an efficient 
mathematical shortcut and visualization for implementing it.  
We then turn to state functions representing systematic uncertainties, 
whose evolution can cause incorrect cosmological conclusions.  
In Sec.~\ref{sec:pop} we consider supernova population fractions as the 
constrained state function and investigate the biases 
this can impose on cosmological parameters, and how to best constrain 
these with redshift specific observations.  We discuss dust extinction 
corrections and their interaction with filter calibration errors in 
Sec.~\ref{sec:dust}, and how to control these with wavelength specific 
measurements.  The summary is presented in Sec.~\ref{sec:concl}.

\section{Realizations of the Equation of State \label{sec:method}} 

We begin by phrasing the analysis in terms of the dark energy equation of 
state, although the results are generally applicable to any state function. 

One possible goal for propagating an array of equation of state functions to 
observational constraints is to place as little prior constraint on 
the functions as possible, an admission of maximal ignorance in the 
hope that the observations impose form on chaos.  A more restrained 
approach is to treat the form of {\it deviations\/} from the basic 
function as free, perhaps representing unknown systematic uncertainties, 
though bounded in amplitude in some way.  The most direct approach then 
is to describe the deviations $w(z)-w_b(z)$ by some value in each small 
redshift bin, equivalent to expanding in a top hat basis. 

This can be transformed into any other orthogonal basis and we can 
hope that a principal component analysis lets us compress the information 
in some way, such that a small, tractable number of modes gives a 
simplified, though still somewhat diverse, functional form.  
We can write 
\beq 
w(z)-w_b(z)=\sum_i \alpha_i\,e_i(z)\,, 
\eeq 
where we refer to $w(z)-w_b(z)$ as the state function, $e_i(z)$ as 
the modes or principal components, and $\alpha_i$ as the mode 
coefficients.  Note that the state function is really the deviation 
from some baseline, and can represent the dark energy equation of 
state or the cosmic ionization fraction, supernova subclass population 
fraction, etc.  In the top hat basis, $e_i(z)$ would simply be 1 within 
the appropriate redshift bin and 0 outside, and $\alpha_i$ would simply 
be $w_i$, the value of the state function within the bin. 

The state function may not be allowed to have arbitrary excursions, 
but can be constrained by physical or theoretical expectations to 
lie within some bounds. 
These bounds could be elementary, such as the ionization fraction 
must lie between 0 and 1, or more physical, such as the equation of 
state for a minimally coupled, canonical scalar field must possess 
$w(z)\in[-1,+1]$.  We define the bounding function, or envelope, by 
\beq 
\wm(z)\le w(z)-w_b(z) \le \wpcap(z) \,. 
\eeq 

Given real data, the results should localize within the bounds.  One 
might be tempted to loosen the bounds and allow the data to lead to 
the proper area of parameter space.  However, the data does not always 
have the required leverage to make this a successful approach.  For 
example, if the equation of state rapidly oscillated between $-10$ and 
$+10$, this could not be detected in the distance measurements (or 
such an oscillation in ionization fraction, even to unphysical negative 
values, might not be seen in cosmic microwave background 
polarization measurements) but one 
has spent a lot of effort calculating over an enlarged range.  
Furthermore, 
when dealing with systematics, unknown by definition, or projected 
future measurements, if one does not bound the amplitudes then no real 
information can be obtained from the results.  Thus, one has to balance 
reasonable, physical bounds and the computational efficiency with the 
desire not to restrict the input.  We take $\wm$, $\wpcap$ to be defined 
with this in mind. 
The effects on the principal components of increasing the envelope are 
simply given by scaling $\wm$, $\wpcap$ in the formulas derived.  

The question then becomes how to best incorporate these bounds in 
``configuration'' (e.g.\ redshift) space 
into the coefficients of the principal components (PC) in mode space. 
For example, to generate realizations of state functions 
that are viable according to the bounds imposed, we must know how to 
properly sample the PC coefficients. 

Several methods can be attempted, but must be assessed for their 
(computational) efficiency, purity, and completeness.  The obvious, 
and least efficient method is simply to try values of the coefficients 
$\{\alpha_1,\dots,\alpha_N\}$ and see if the functions $w(z)$ obey 
the bounds at each redshift.  In practice one must truncate the 
number of PCs at a finite number, choose a finite range for each 
$\alpha_i$, and with a number of grid points $R$ sampling the coefficient range 
evaluate $R^N$ functions to test whether they lie within the bounds. 
For a grid of 20 points and 10 PCs, this requires $20^{10}\approx 10^{13}$ 
evaluations.  We call this the scanning strategy.  It would be pure 
and complete, but is not efficient. 

A second approach is to ask that each PC contribution to the state 
function obey the bounds individually.  This mode-by-mode strategy 
has been implemented in \cite{mort1,tobin,mhh} for example.  Projecting 
a given PC against the state function yields the coefficient: 
\beq 
\alpha_i=\caln\int dz\,e_i(z)\,[w(z)-w_b(z)]\,, 
\eeq 
where $\caln=1/\int dz\, e_i^2(z)$ is the normalization factor. 
Incorporating the bounds on the state function and breaking the 
integration region into those redshifts where $e_i(z)$ is positive 
and those where it is negative, one obtains the bounds 
\beq 
\alpha_i^-\le \alpha_i\le \alpha_i^+\,, \label{eq:modelim} 
\eeq 
where 
\beqa 
\alpha_i^\pm=(\caln/2)\int dz &\,\,\bigl\{\,\,&[\wpcap(z)+\wm(z)]\,e_i(z) \nonumber\\ 
&\pm&[\wpcap(z)-\wm(z)]\,|e_i(z)|\bigr\}\,. \label{eq:alfbound} 
\eeqa 

The main problem with this approach is that the modes are treated 
independently.  So if one saturates the bounds on each coefficient, 
say, then the generated state function may actually lie outside the 
envelope.  
Thus, this method is complete but not pure. 

One way to incorporate all the mode information is to consider the 
integral of the square of the state function \cite{mhh}.  Then 
\beq 
\int dz\,[w(z)-w_b(z)]^2=\caln\sum_i \alpha_i^2\,. 
\eeq 
Imposing the bounds on the state function then delivers the constraint 
\beq 
\sum_i \alpha_i^2\le (1/\caln) \int dz\, {\rm max}\{\wpcap^2(z),\wm^2(z)\}\,,  \label{eq:spherelim} 
\eeq 
where the maximum is to be evaluated for each redshift. 
(This generalizes the expression in \cite{mhh} to when the envelope 
is not redshift independent.)  We call this the integrated method and it 
defines a sphere in the mode coefficient space.  This approach guarantees 
completeness but not purity, i.e.\ every viable state function can 
be generated with this set of coefficients, but nonviable ones can be 
as well.  By itself it lacks a specific prescription for implementing the selection of $\alpha_i$'s. 

An alternate approach is the ``global'' method, where coefficients 
are chosen based on the previous coefficient values.  For example, 
choose the coefficient $\alpha_1$ based on the envelope constraint 
as if this were the only mode, giving 
\beq 
\frac{W_\mp(z)}{e_1(z)}\le\alpha_1\le\frac{W_\pm(z)}{e_1(z)}\,, \label{eq:alfglob} 
\eeq 
where the top (bottom) sign holds for $e_1(z)>0$ ($<0$).  This is 
applied for all redshifts under consideration and the tightest 
constraints obtained define the range of $\alpha_1$.  Once an $\alpha_1$ 
is sampled within the allowed range, one obtains similar bounds on 
$\alpha_2$ using, e.g., $\alpha_2\,e_2(z)\le\wpcap(z)-\alpha_1\,e_1(z)$, 
and so on.  This global method is simpler, not involving any integrals, 
although it still involves scanning over choices for the coefficients 
within their allowed range.  
However, a value for the coefficient $\alpha_i$ that has been rejected 
because it lies outside the bounds of Eq.~(\ref{eq:alfglob}) or similar 
may actually be valid because another PC counteracts its contribution 
and pulls the state function back within the envelope.  Thus the method 
is pure, i.e.\ all generated state functions will be viable, but not 
complete.  

Thus we have generating methods that are pure and complete but 
inefficient (scanning method), complete (mode-by-mode method), and 
pure (global method), but no obviously optimal method.  We address 
this lack in the next section, and show how all the methods are related.

\section{A Pure and Complete Prescription \label{sec:hypercube}} 

The physical bounds on the state function are imposed in the 
redshift space but we need to translate these into PC coefficient 
space if we want to generate principal component analysis (PCA) realizations of the state function.  
The problem is that a principal component contributes to the state 
function over the whole redshift interval considered.  Effectively, PCA 
mixes the values $w_i$ from all redshift bins in the bin basis.  
Therefore what the 
chosen bound corresponds to -- a value for the state function at a 
particular redshift $z$ -- is not localized in coefficient parameter 
space but is described by a linear combination of many modes weighted 
by their respective coefficients.  

\subsection{Hypersurface Picture} 

However, by making use of the 
properties of linear transformations, and a particularly clear geometric 
picture, we can implement an exact, fast method for the translation. 
Consider the envelope on a single $w_i$.  This gives a range, or line 
segment, along the $w_i$ axis.  Combining the envelopes for all redshift 
bins, i.e.\ $w_i$ parameters, defines a hypersurface in an $N$ 
dimensional space, where $N$ is the number of redshift bins.  If the 
bin bounds do not depend on values $w_i$ in 
other bins, i.e.\ each bin is independent (recall the original motivation 
was to consider state function behaviors without assuming a functional 
form), then the surface is a hyperrectangle.  

The corners of the hyperrectangle are defined by the values $W_\pm(z_i)$ 
of the envelope.  These $2^N$ vertices contain all information on the 
boundary between the permitted, i.e.\ viable, instances of state functions 
$w(z)-w_b(z)$ and the disallowed or unviable ones.  That is, the 
boundary defines the pure and complete set.  

\begin{figure}[h]
\begin{centering}
\includegraphics[width=\columnwidth]{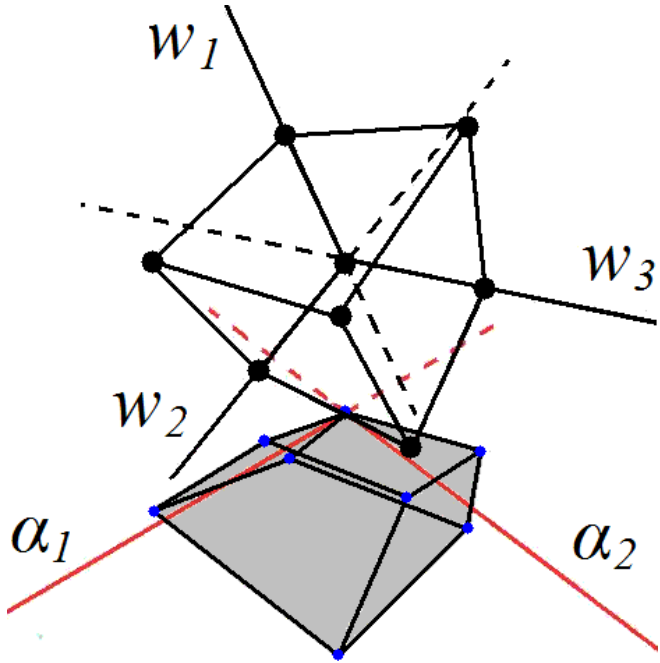}
\end{centering}
\caption{A cube in bin space, corresponding to redshift independent 
limits on $w(z)-w_b(z)$ in 3 redshift bins (i.e.\ the simplest 
nontrivial case of an $N$-dimensional hyperrectangle), is projected onto 
the parameter space of the coefficients of the first two principal 
components.  The small blue dots indicate the projection of the cube 
vertices.  The permitted area in the $\alpha_1$-$\alpha_2$ plane is 
indicated by the light grey shading.  Note that it does not form a 
rectangle.} 
\label{fig:cube} 
\end{figure}

We defined the vertices as sets of $w_{i}$ coordinates but now let us 
consider the hypersurface in the PC coefficient space of $\alpha_i$ 
coordinates.  Because the PCA is a (normalized) linear transformation of the redshift 
bin values, the hypersurface is merely rotated, not distorted or expanded. 
If we are interested in a subset of $M$ modes, smaller than the maximum 
number $N$ (there cannot be 
more modes than the original bins used to define the PCs), then this 
corresponds simply to a projection of the hypersurface onto the subspace 
of the $M$ PC coefficients.  We illustrate the case of a 3 dimensional 
hyperrectangle projected onto 2 PC coefficients in Figure~\ref{fig:cube}.

The boundary of the pure and complete set of PC coefficients is 
defined by connecting the outermost projected vertices ensuring the 
boundary remains convex.  This follows from the linearity of the 
transform: the hypersurface is convex and so the projection must then 
itself be convex.  (Note the projected figure is not in general 
rectangular.)  The projection can be computed quite quickly 
through the use of matrix algebra (see the Appendix).  Thus a pure 
and complete set of PC coefficients for viable, and only viable, 
state functions can be generated efficiently. 

If we do a complete projection over all $N$ dimensions except one, 
say $\alpha_j$, we obtain the absolute minimum and maximum bounds on 
$\alpha_j$.  These bounds are equivalent to those found with the 
mode-by-mode method, Eq.~(\ref{eq:modelim}).  Furthermore, since 
distances are conserved under the linear transformation, the 
maximum distance in bin space, i.e.\ the longest diagonal of the 
hyperrectangle, must also be the maximum distance in PC coefficient 
space.  A (hyper)sphere with this diameter circumscribes the allowed 
set and corresponds to the sphere of the integrated method, 
Eq.~(\ref{eq:spherelim}).  From the fact that the hypersphere 
circumscribes the hyperrectangle, it is clear that this method 
generates a complete, but not pure, set. 

The degree of impurity or incompleteness for the methods can be tied 
to the ratios of areas (or hypervolumes) between the geometric figure 
defined by the methods and the true hyperrectangle.  If the PC 
coefficients are highly independent of each other (of course the mode 
vectors $e_i$ themselves are orthogonal), then we expect the mode-by-mode 
method, where we ignored the effect of $\alpha_j$ on $\alpha_i$, to be 
a good approximation, i.e.\ nearly pure and complete.  Taking zero 
correlation between coefficients defines a rectangle in the 
$\alpha_i$-$\alpha_j$ plane, for any $\alpha_i$, $\alpha_j$.  As the 
coefficient parameters become more correlated, the mode-by-mode method 
should become less efficient at finding only the viable state 
functions, i.e.\ less pure.  Geometrically, the filling factor of the 
true hyperrectangle projection will decrease. 

\begin{figure}[!h]
\begin{centering}
\includegraphics[width=\columnwidth]{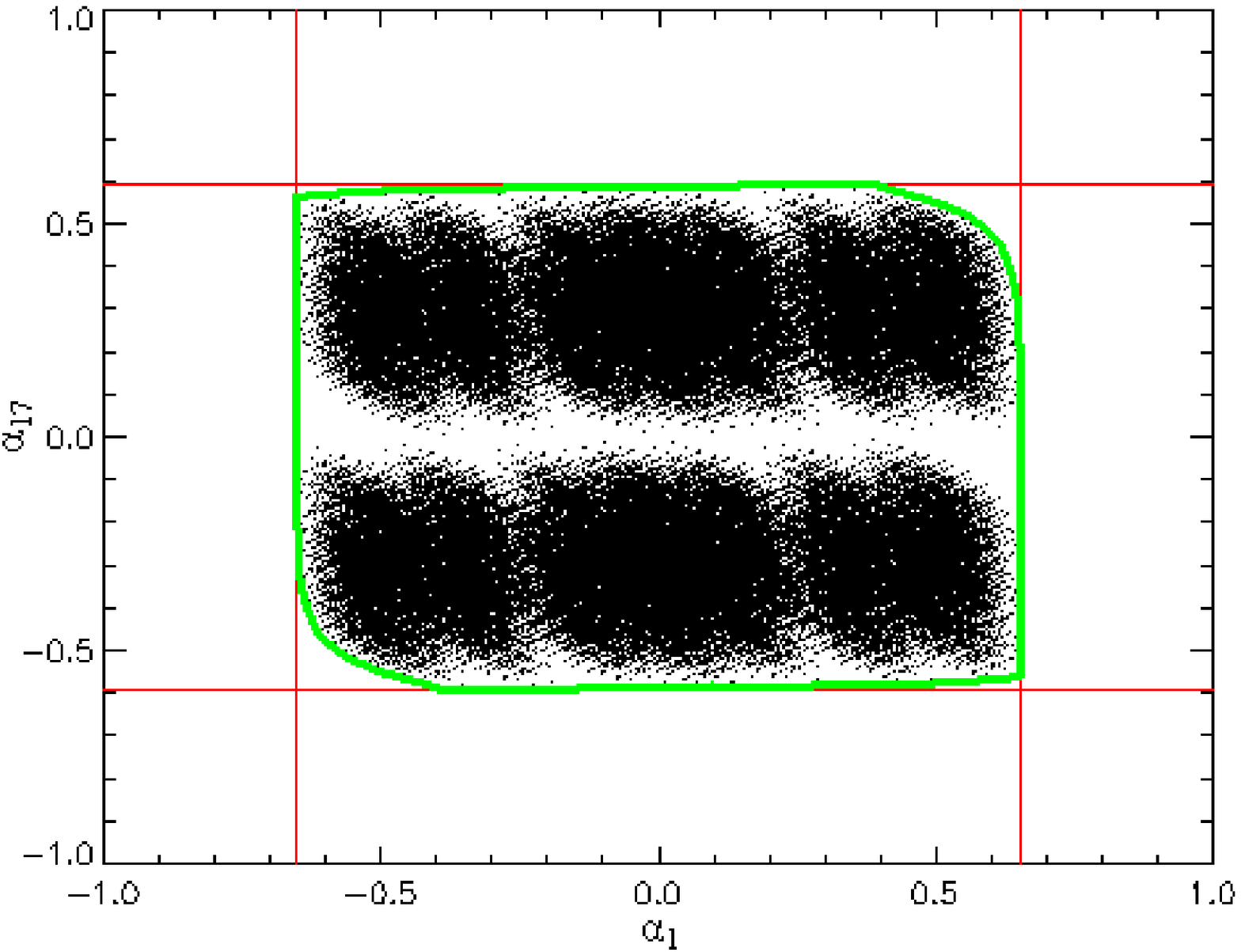}
\includegraphics[width=\columnwidth]{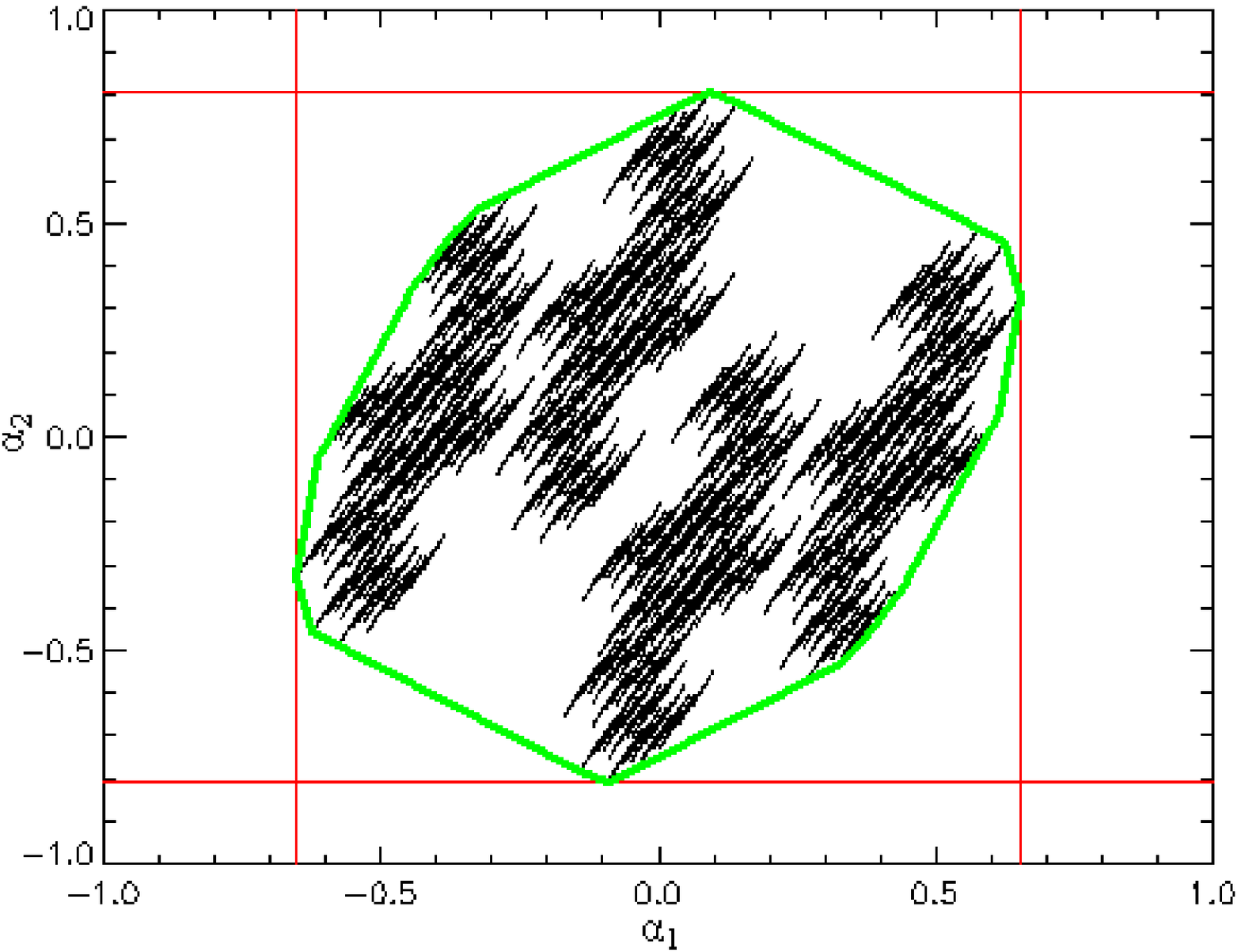}
\end{centering}
\caption{Projections onto PCA coefficient planes are shown for the 17 
mode case.  The light, red vertical and horizontal lines indicate the 
mode-by-mode bounds of Eq.~(\ref{eq:modelim}).  The dots give 
the projection of the $2^{17}$ vertices of the hyperrectangle, and the 
thick, green polygon gives the outer boundary, defining the interior 
region of pure and complete viable state functions.  [Top panel] The 
$\alpha_1$-$\alpha_{17}$ plane in parameter space has low correlation 
between these modes, so the filling factor of the approximate bounds 
is high.  [Bottom panel] The 
$\alpha_1$-$\alpha_2$ plane has strong correlations because the PCs 
overlap substantially in redshift, so the filling factor is low. } 
\label{fig:10D} 
\end{figure}

Figure~\ref{fig:10D} illustrates this relation between correlation and 
filling factor.  The top panel shows the projection onto the space 
spanned by the coefficients of PC modes 1 and 17.  Since $e_1$ and 
$e_{17}$ have their main weights at very different redshifts, the 
coefficients $\alpha_1$ and $\alpha_{17}$ are substantially 
uncorrelated, and indeed the filling factor is high (but not perfect). 
The bottom panel displays the equivalent projection for modes 1 and 2. 
Here the overlap of the modes in redshift is greater and so the 
coefficients are more correlated; the filling factor is noticeably 
decreased.  Therefore the mode-by-mode method is not efficient when 
considering the dominant modes.

By contrast, the exact hyperrectangle projection method is highly 
efficient.  For 10 modes, say, there are $2^{10}\approx 10^3$ vertices 
to evaluate (the projection takes negligible computational time using 
the method in the Appendix).  Contrast this with the previous $10^{13}$ 
evaluations needed for direct scanning. 

If we were to increase the number of redshift bins (i.e.\ bin modes, 
holding the 
redshift range constant), this allows for more and more PC modes. 
However, since most of these additional modes would be less and less 
correlated with a given mode, we effectively have a convergence in 
the behavior of the parameters, i.e.\ the projected boundary in a 
given $\alpha_i$-$\alpha_j$ plane.  

\begin{figure}[!h]
\begin{centering}
\includegraphics[width=\columnwidth]{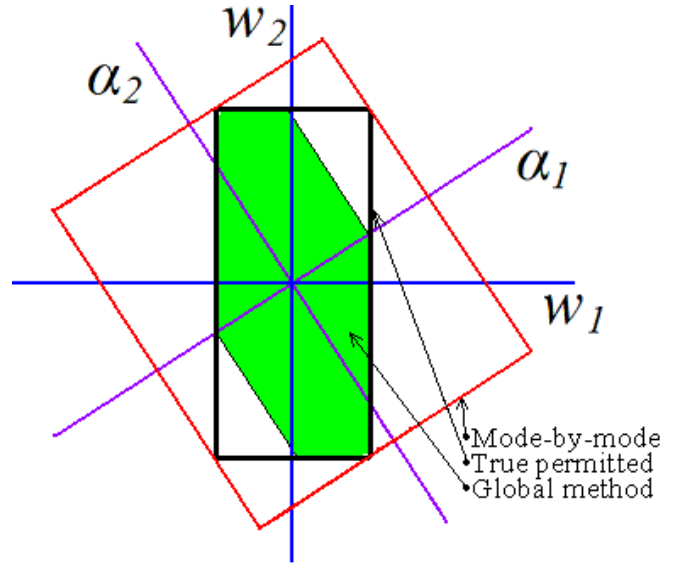}
\end{centering}
\caption{Three different approaches to obtaining the principal component 
coefficients for a constrained state function are illustrated, here for 
the first and second coefficients and the simplest case with only 2 modes.  
The true viable region is the thick black rectangle in $w_1$-$w_2$.  
The shaded, green interior polygon shows the pure but incomplete global 
approximation, while the exterior thin, red rectangle shows the 
complete but impure mode-by-mode approximation.  We do not show the 
circle circumscribing the outer rectangle that corresponds to 
the complete but impure integrated approximation.} 
\label{fig:compare22}
\end{figure}

\begin{figure}[!h]
\begin{centering}
\includegraphics[width=\columnwidth]{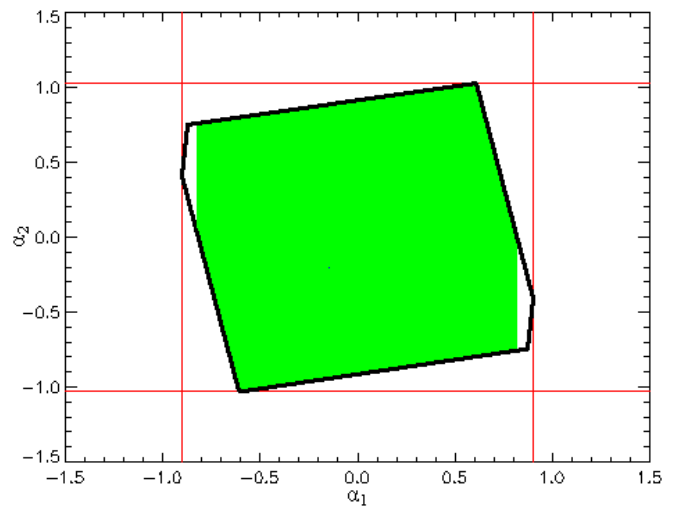}
\end{centering}
\caption{As Fig.~\ref{fig:compare22} but for 3 binned $w$ modes projected 
onto the plane of the first two PC coefficients.  Here we 
suppress the axes.  The light, red vertical and horizontal lines 
show the mode-by-mode limits; the thick, black polygon defines the 
exact, pure and complete function space, and the green, shaded region 
is the global approximation.} 
\label{fig:compare32}
\end{figure}

In summary, we have presented an efficient, pure, and complete method of 
obtaining the boundary defining the set of viable state functions.  The 
relation to previous (not simultaneously pure and complete) methods is 
illustrated in Figs.~\ref{fig:compare22}-\ref{fig:compare32}.  The outer 
rectangle gives the prescription of the mode-by-mode approach; the thick 
interior polygon shows the exact solution using the projection of the 
hyperrectangle; and the light shaded interior non-rectangle illustrates 
the global method, representing a cut through the hyperrectangle at 
$\alpha_{i>2}=0$. Since the exact solution can be generated efficiently 
there is no need to use the over- (global) or under- (mode-by-mode) 
approximation.

While we have solved the problem of obtaining efficiently the constraint 
on the region of principal component space that is viable given some 
bounds on the state function, we have to ask whether this is really the 
best path for analyzing the effect of various state functions. 
To scan over all viable PCs we would select from the PC coefficients within 
the allowed region.  If the probability of the state function in the bin 
basis was uniform within the bounds, then because of the linearity of 
the principal component transformation the interior volume in PC coefficient 
space can also be uniformly sampled.  However, in general we would have 
some correlation 
\beq 
\langle \alpha_i \alpha_j\rangle=e_{ip}\,e_{jq}\,\langle w_p w_q\rangle\,, 
\eeq 
where angle brackets denote the ensemble average and $p$, $q$ are 
redshift bin indices (implicitly summed) while $i$, $j$ are component 
mode indices. 

Writing this in matrix notation, 
\beq 
A=EWE^T\,, 
\eeq 
where $W$ is the correlation of the state function (e.g.\ equation of 
state values in redshift bins) and $A$ is the correlation that then must 
be imposed on the selection of PC coefficients.  Note that when the 
bounds on the state function are redshift dependent -- as when some 
data constraint knowledge is incorporated -- then even a diagonal $W$ does not 
lead to a diagonal $A$.  

\subsection{Restricting Modes} 

\begin{figure}[!h]
\begin{centering}
\includegraphics[width=\columnwidth]{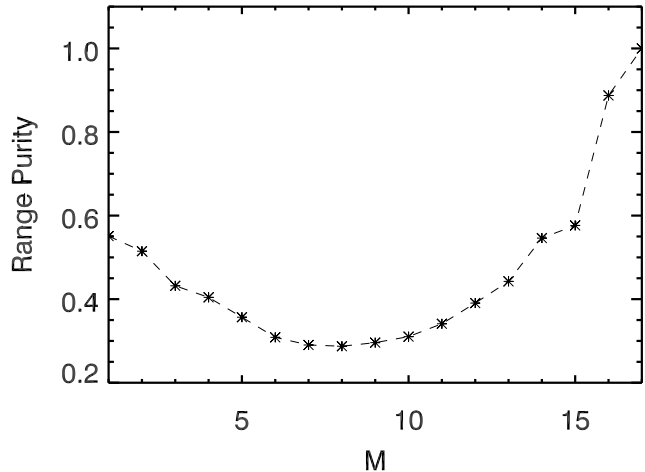}
\end{centering}
\caption{When not all PC modes are kept then the purity of the 
mode-by-mode method decreases. 
The curve shows the range purity -- the product of the bounds on each 
PC coefficient for the exact method when only $M$ modes are retained 
vs.\ when all modes are.   This is also equivalent to the ratio of 
the $M$ mode range area to the mode-by-mode method's area.} 
\label{fig:spaceratio}
\end{figure}

If we keep only $M$ modes in PC space then, because the 
$\alpha_i$ axes are not in general aligned with the bin basis $w_i$ 
axes, the PCs will still span the redshift range but will not be able 
to describe the full range of $w(z)$ behaviors within the true bounds. 
That is, we diminish the completeness if we restrict the number of PC 
modes.  Note this can be treated in the hyperrectangle 
picture as slices through the $N$-rectangle at fixed values of the 
neglected $N-M$ parameters (see the Appendix for more details). 
Figure~\ref{fig:subspace} shows an example of the diminished state 
function space accessed when limiting to 4 modes (out of 17).  Moreover, 
the impurity of the mode-by-mode method becomes more severe, with 
Fig.~\ref{fig:spaceratio} showing that using less than the full number 
of modes can yield up to 70\% of the generated forms of the state 
function being spurious, i.e.\ ones that invalidly exceed the bounds,  
under the restriction to the first $M$ modes.

\begin{figure}[!h]
\begin{centering}
\includegraphics[width=\columnwidth]{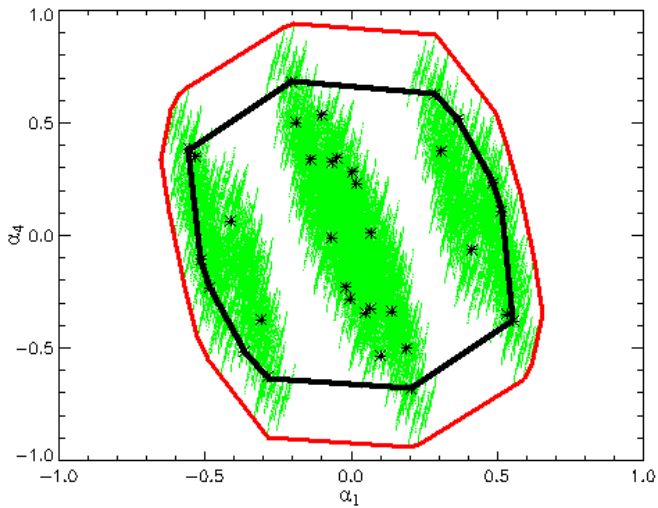}
\end{centering}
\caption{Restricting the number of PC modes kept in the analysis 
will lose completeness in the state function forms allowed.  The 
light, green dots show the projection of the full set of hyperrectangle 
corners, with the light, red outer polygon giving the pure and complete 
bounds in $\alpha_1$-$\alpha_4$ space.  The black dots and black polygon 
show the case when only the first 4 modes are kept.  The region in between 
the polygons represents viable, but lost state functions.} 
\label{fig:subspace}
\end{figure}

In the end, then, because of the coefficient correlations and the 
completeness issues, little advantage accrues in fact to the use of PCA 
for the scanning over functional forms. 
It is more efficient (and innately pure and complete) simply to carry 
out the analysis in the original 
state function space where the constraints originated.  A standard 
redshift bin basis allows the freedom needed to model the form of 
the state function, and the constraints can be imposed naturally 
without complicating the generation of realizations. 
In the next sections we demonstrate the real world application of the 
bin basis state functions to problems involving calculating the effects 
on cosmology results when confronted with unknown systematics functions.

\section{Systematics: Supernova Population Drift \label{sec:pop}} 

The use of constrained functions, and their impact on parameter 
estimation or the science results, enters into myriad areas of 
cosmology.  This is a particularly important issue for systematic 
uncertainties, where we do not know the form of the residual error 
function.  We therefore consider the example of the population fraction of a 
certain type of source as a key element of the cosmology calculation 
and take as the state function the uncertainty in our knowledge of 
it.  By definition the function is constrained to take 
values in the range $[0,1]$.  If the source is a standardized distance 
indicator such as Type Ia supernovae (SN) and we posit that the populations 
represent subclasses with slightly different intrinsic magnitudes, then 
any variation with redshift in the population fractions will appear as 
magnitude evolution and, if unrecognized, bias the cosmological parameter 
estimation.  This is known as population drift (for theoretical 
discussion and observational limits see 
\cite{coping,sullivan,conley07,bronder,howell08,sullivan09}). 

In \cite{lin0812}, the effects of population drift as a bias or increased 
dispersion (if adding fit parameters) on cosmology were investigated for 
a class of state functions depending as a power law in redshift (also 
see \cite{virey}).  
Here we can analyze {\it every\/} form of population drift and 
investigate which are the most dangerous.  In addition we refine our 
quantification of the cosmology bias and explore in what redshift ranges 
the population drift systematic is most biasing. 

Population drift as a systematic relies on two elements: an actual 
difference in intrinsic magnitudes between the subclasses and a redshift 
dependence in the difference.  A mere constant difference is absorbed 
into the absolute 
magnitude nuisance parameter $\mathcal{M}$.  For simplicity, we illustrate 
the basic results for a two population model, where the SN have a 
fraction $f_1(z)$ with intrinsic magnitude $M_1$ and a fraction $f_2(z)$ 
with intrinsic magnitude $M_2$.  We can consider $f_1$ as representing 
all the populations we recognize and $f_2=1-f_1$ as an aggregate of those 
unrecognized.  Then the unrecognized systematics appears as a magnitude 
evolution  
\beq 
\Delta m(z)=\Delta M\,[f(z)-f(0)]\,, \label{eq:dmpop} 
\eeq 
where $\Delta M=M_2-M_1$ and $f=f_2$ is now our state function.  The 
constraint on the state function, by definition of the population 
fraction, is $f\in[0,1]$. 

Propagating this systematic through to the cosmology parameters is 
straightforward.  For a parameter set $\vec p=\{p_i\}$, the bias is 
(see, e.g., \cite{linbias}) 
\beq 
\delta\vec p=(U^{T}C^{-1}U)^{-1}\,U^{T}C^{-1}\,\Delta O\,, \label{eq:biasgen} 
\eeq 
where $O$ is the observable, $U=\partial O/\partial\vec p$, $C$ the error 
covariance matrix for the observables and $\Delta O$ is a systematic offset 
in observable $O$. The term in parentheses is simply the Fisher matrix, 
and so its inverse is the parameter covariance matrix.  In the case we 
are currently considering the observables $O$ are SN magnitudes at various  
redshifts and $\Delta O$ is the magnitude offset of Eq.~(\ref{eq:dmpop}). 
For a diagonal error covariance matrix the equation takes a simpler form 
\beq 
\delta p_{i}=\Delta M\,(F^{-1})_{ij}\sum_{k=1}^{N}f(z_k) 
\frac{1}{\sigma^{2}(m_{k})}\frac{\partial m_{k}}{\partial p_{j}}\,. \label{eq:biasm} 
\eeq 
Here $F$ is the Fisher matrix, say $4\times4$ with respect to $\mathcal{M}$, 
$\Omega_{m}$, $w_{0}$, and $w_{a}$, where $\om$ is the present matter 
density in units of the critical density.  There are $N$ data points, 
each with an associated redshift $z_k$. 

We can now explore the effect of any form for $f(z)$, subject only to 
the constraint $0\le f(z)\le1$.  We do not need to assume a functional 
form for $f$, rather we want to allow it complete freedom under constraint. 
As we saw in the previous section, no real advantage accrues to a principal 
component analysis -- all the information exists and is more accessible 
using a redshift bin basis.  In fact, PCA when keeping only a more limited 
number of modes loses information and the bin basis allows for greater 
efficiency in scanning the allowed state function parameter space. 

Before we calculate the bias we examine in more detail how to assess it 
quantitatively.  The bias $\delta p_i$ itself is only informative together 
with the cosmological parameter uncertainties.  If the estimated 
uncertainty on the parameters is large, then the relative effect of a 
particular bias is lower, meaning that the (mis)estimated model is still 
within some acceptable confidence level contour.  For each parameter 
$p_i$, \cite{lin0812} employed the risk statistic \cite{kendall} 
\beq 
{\rm Risk}(p_i)=\sqrt{\sigma_{p_i}^2+\delta p_i^2} 
\eeq 
as a measure of the influence of the bias.  However, the overall 
cosmology is biased by the vector $\delta\vec p$.  One could imagine that each parameter 
bias relative to the dispersion is small, but in a direction such that 
$\delta\vec p$ is oriented along the thin part (minor axis) of the confidence 
level contour; then  a small shift could actually be a large bias 
relative to the contour, i.e.\ in terms of the $\dcc$.  Following 
\cite{shapiro} (cf.\ \cite{dodelson}) therefore, we use as our bias statistic 
\beq 
\dcc=\boldsymbol{\delta p}\,\mathbf{F}^{(r)}\,\boldsymbol{\delta p}^T\,,\label{eq:dchi} 
\eeq 
where $\boldsymbol{\delta p}$ is the vector of parameter biases we consider 
and $\mathbf{F}^{(r)}$ is the reduced Fisher matrix, marginalized over all 
parameters except those in whose biases we are interested.  For example, 
if we consider biases in the $w_0$-$w_a$ contour, then $\mathbf{F}^{(r)}$ 
is the inverse of the $2\times2$ submatrix of the covariance matrix 
containing $w_0$ and 
$w_a$.  In the case of a single parameter, $\dcc=(\delta p/\sigma_p)^2$ 
and Risk $=\sigma_p\sqrt{1+\dcc}$. 

We can now scan over all possible population drifts $f(z)-f(0)$ and 
evaluate the bias effects on the cosmological parameters.  We write 
$f$ in the redshift bin basis, initially with 17 bins uniform 
between $z=0-1.7$. 
For the Fisher matrix we take simulated data based on the SNAP SN 
redshift and error distribution \cite{snap}, 
plus a Planck-inspired constraint on the reduced distance to CMB last 
scattering of 0.2\%.  The fiducial cosmology is $\Lambda$CDM with 
matter density $\om=0.28$. 

Table~\ref{tab:poptrs} describes the population evolution functions 
computed to deliver the maximum bias in $\dcc$.  The results have a very 
simple form: a single or double sharp transition in redshift.  This 
can be understood through analyzing Eq.~(\ref{eq:biasm}).  The bias 
is a linear transformation of $f$, hence for a maximum bias $f$ is 
driven to the extreme value that complements the sign of the term 
$\beta_i\equiv(F^{-1})_{ij}\, \partial m_k/\partial p_j$, for each $z_k$.  
To maximize $\delta p_i$, when $\beta_i>0$ 
then $f$ should be 1, while when $\beta_i<0$ then $f$ should be 0.  
This will give coherent addition of the terms in the sum and so deliver 
the largest $\delta p_i$.  Thus $f$ should simply be a series of tophats 
over those redshifts where $\beta_i(z_k)$ is positive.  For the parameter 
$\om$, $\beta_{\Omega_m}$ crosses once through 0 so $\delta\om$ is 
maximized by a population function that has a single transition, at 
$z\approx0.7$.  Thus the most potent evolution function $f$ -- the one 
having the strongest consequence for cosmology estimation -- has a 
step appearing at $z=0.7$ and extending to the maximum redshift.  
For $w_0$ or $w_a$, the respective $\beta$'s cross twice through 0 so 
$f$ forms a tophat extending from $z\approx0.2$ (respectively 0.1) to 
$z\approx1.0$.

\begin{table}[h]
\begin{tabular}{cccc}
\hline 
Parameters \quad & $z_{\rm trans}$ \quad & max $\Delta\chi^{2}_{\Delta 
M=0.01}$ \quad & $\Delta M\,(1\,\sigma)$ \\ 
\hline 
$\Omega_{m}$ & 0.7 & $0.92$ & $0.010$ \\ 
$w_{0}$ & 0.2, 1.0 & $1.35$ & $0.0086$ \\ 
$w_{a}$ & 0.1, 1.0 & $1.35$ & $0.0086$ \\ 
$w_{0},w_{a}$ & 0.1, 1.0 & $1.36$ & $0.013$ \\ 
$\Omega_{m},w_{0},w_{a}$ & 0.1, 0.9 & $1.39$ & $0.016$ \\ 
\hline 
\end{tabular} 
\caption{For each set of parameters we consider the form of population 
evolution that maximizes the cosmology bias in terms of $\dcc$.  The 
$z_{\rm trans}$ column gives the redshift of the maximizing step function 
in $f$, delivering a maximum bias $\dcc$ scaled to the case where 
$\Delta M=0.01$, shown in the next column.  Note $\dcc$ will scale as 
$\Delta M^2$.  The last 
column shows the value of $\Delta M$ that will shift the derived 
cosmology by $1\sigma$ from the true cosmology.} 
\label{tab:poptrs} 
\end{table}

This has a number of crucial implications.  First, since the state function 
with the maximal effect arises from a sharp transition, we see that it 
was prescient to use the bin basis for the population function.  Had we 
transformed to principal component space, or some smooth orthogonal basis 
such as the Chebyshev polynomials considered by \cite{lin0812,amara}, then 
we would have had difficulty approximating the true solution with a finite 
number of modes.  Second, the sensitivity to population evolution at 
specific redshifts guides the survey design to obtain especially detailed 
measurements at these redshifts.  The results indicate that as observations 
make the transition from local ($z<0.1$) SN to low redshift ($z\gtrsim0.2$) 
SN, they must comprehensively collect and study the SN properties so as to 
ensure a firm like-to-like comparison and not allow for unrecognized 
populations.  Similarly, the transition from $z<1$ to $z>1$ SN is key, 
so a transition from ground-based observing to the space-based observing 
necessitated at $z>1$ could be problematic.  A homogeneous survey 
extending across this transition would have far better control over the 
systematic uncertainty. 

These results hold as well when considering simultaneously bias in 
multiple parameters in the $\dcc$ formalism.  The population $f$ enters 
Eq.~(\ref{eq:dchi}) quadratically and so the bias is still maximized by 
the extreme values of $f$, i.e.\ top hats in redshift.  The transition 
locations do not shift appreciably when considering bias in the two 
parameter space of $w_0$-$w_a$ nor the three parameters $\om$, $w_0$, $w_a$ 
simultaneously.  Furthermore, the transition locations are robust to 
changing the step functions to more gradual slopes. 

The last column of Table~\ref{tab:poptrs} shows the magnitude of $\Delta M$ 
that in the worst case of population evolution causes a $1\sigma$ 
misestimation of the cosmology.   This is where the scaling of the state 
function bounds enters: the shape, i.e.\ redshift dependence of the 
population function, is unaffected by amplitude of the bounds, but the 
absolute level is determined by the bounds.  If we consider twice as large 
values for $\Delta M$ (or if we were to unphysically allow $f$ to range 
from 0 to 2), then $\Delta\chi^2$ just scales with $\Delta M^2$.  
If we want to be sure that population 
drift cannot cause a $>1\sigma$ shift in the equation of state parameters, 
we need to be able to recognize SN subclasses differing by 
0.0086 mag or more. 

\begin{figure}
\begin{centering}
\includegraphics[width=\columnwidth]{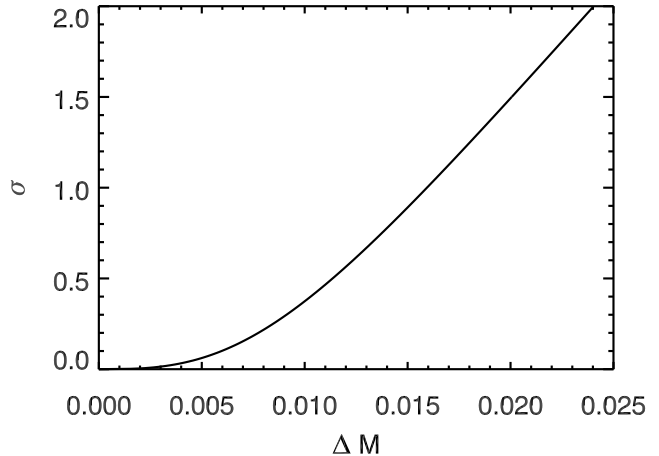} 
\end{centering} 
\caption{For the maximally biasing population drift the cosmological 
parameter set $\{\om,w_o,w_a\}$ is biased by $1\sigma$ for a population 
magnitude difference $\Delta M=0.016$ mag.  The curve shows by how many 
$\sigma$ the best fit cosmology is biased as a function of $\Delta M$.} 
\label{fig:sigM} 
\end{figure}

Figure~\ref{fig:sigM} 
shows the relation between the maximum number of standard deviations 
$\sigma$ by which 
the cosmology is 
distorted, as a function of difference in absolute magnitudes $\Delta M$ 
between the populations, for the full set of cosmological parameters.  
While $\dc^2$ scales as $\Delta M^2$, the number of $\sigma$ this bias 
corresponds to only scales as $\Delta M$ in the one parameter case.  
We see that for three parameters the $\sigma$ remains nearly linear for 
large $\Delta M$ but does not improve as rapidly for $\Delta M<0.015$.

Beyond the maximum $\dcc$, we can investigate other properties of the 
biasing.  For example, we can explore further the direction of the 
systematic shift 
caused in the cosmology parameters, the relation of the forms of 
the population drift, i.e.\ the number of steps or oscillations, to 
the bias, and the overall statistics of the biasing.  

\begin{figure}
\begin{centering}
\includegraphics[width=\columnwidth]{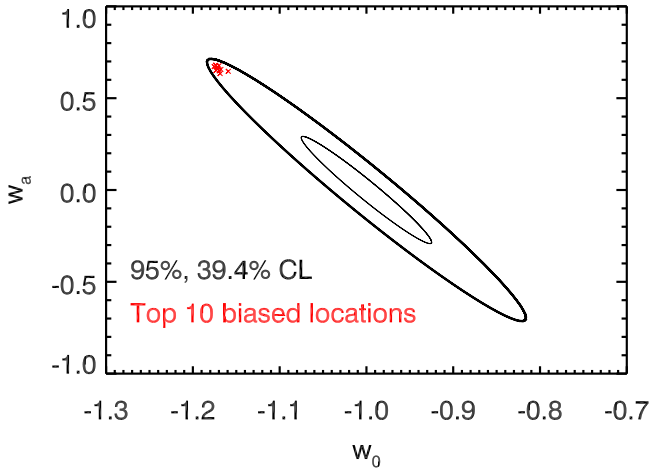}
\end{centering}
\caption{The biases induced for $\Delta M=0.02$ by the worst population 
evolution functions -- those 
that induce the 10 largest $\dcc$ -- are plotted in the $w_0$-$w_a$ plane 
as x's.  Note how they congregate at the extreme end of the major axis 
of the confidence contour.  The inner ellipse indicates the 39\% confidence 
level contour (so $\sigma(w_0)$, $\sigma(w_a)$ are given by direct 
projection to the respective axis), while the outer ellipse shows the 
95\% cl joint likelihood contour.} 
\label{fig:elldc2}
\end{figure}

We begin with 
the effect on the equation of state estimation caused by the systematic 
error.  Figure~\ref{fig:elldc2} shows the specific form of bias induced 
in the equation of state by the 10 worst case population drifts. 
The worst biases all distort the cosmology in the same way: making a 
cosmological constant look like a rapidly varying equation of state. 
Indeed, this is characteristic not just of population drift but of 
any sharp transition in the SN magnitudes, such as from patching 
together two redshift samples with an unrecognized offset (local to 
low redshift samples, or ground-based to space-based).  This points up 
the need for tight crosscalibration, and ideally a continuous, homogeneous 
data set, as well as the need for caution in interpreting 
an apparent behavior of the equation of 
state crossing $w=-1$: exactly what is expected from such a systematic.

\begin{figure}
\begin{centering}
\includegraphics[width=\columnwidth]{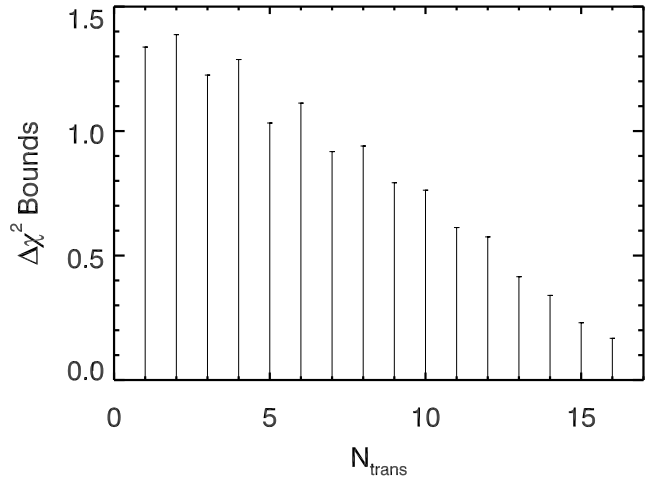}
\end{centering} 
\caption{The range of possible biases $\dcc$ in the cosmology likelihood 
contour for $\om$-$w_0$-$w_a$ is plotted vs.\ the number of transitions 
(changes in value between the 17 redshift bins) in the population evolution.} 
\label{fig:ntrans} 
\end{figure}

The influence of the forms of the population evolution function on the 
bias generated in the cosmology parameters can be investigated through 
looking at the statistics of the $\dcc$ distribution.  For example, 
while the maximum bias is generated from a population function with 
one or two steps at sensitive redshifts, we expect a large number of 
transitions to have relatively little effect since such an oscillatory 
behavior does not resemble the effect of a cosmological parameter.  The 
sum of the terms in Eq.~(\ref{eq:biasm}) effectively cancels out. 
Thus certain types of systematics are fairly benign, such as 
quasi-periodic k-correction errors \cite{daviskim,hsiao}.  
Figure~\ref{fig:ntrans} shows the 
range of $\dcc$ generated as a function 
of the number of transitions in $f$ between redshift bins.  As expected, 
as the number of transitions gets large, the bias decreases.  Similarly, 
when the step amplitude is small then the bias is negligible so the 
$\dcc$ distribution ranges between 0 and the maximum for each number of 
transitions.

\begin{figure}
\begin{centering}
\includegraphics[width=\columnwidth]{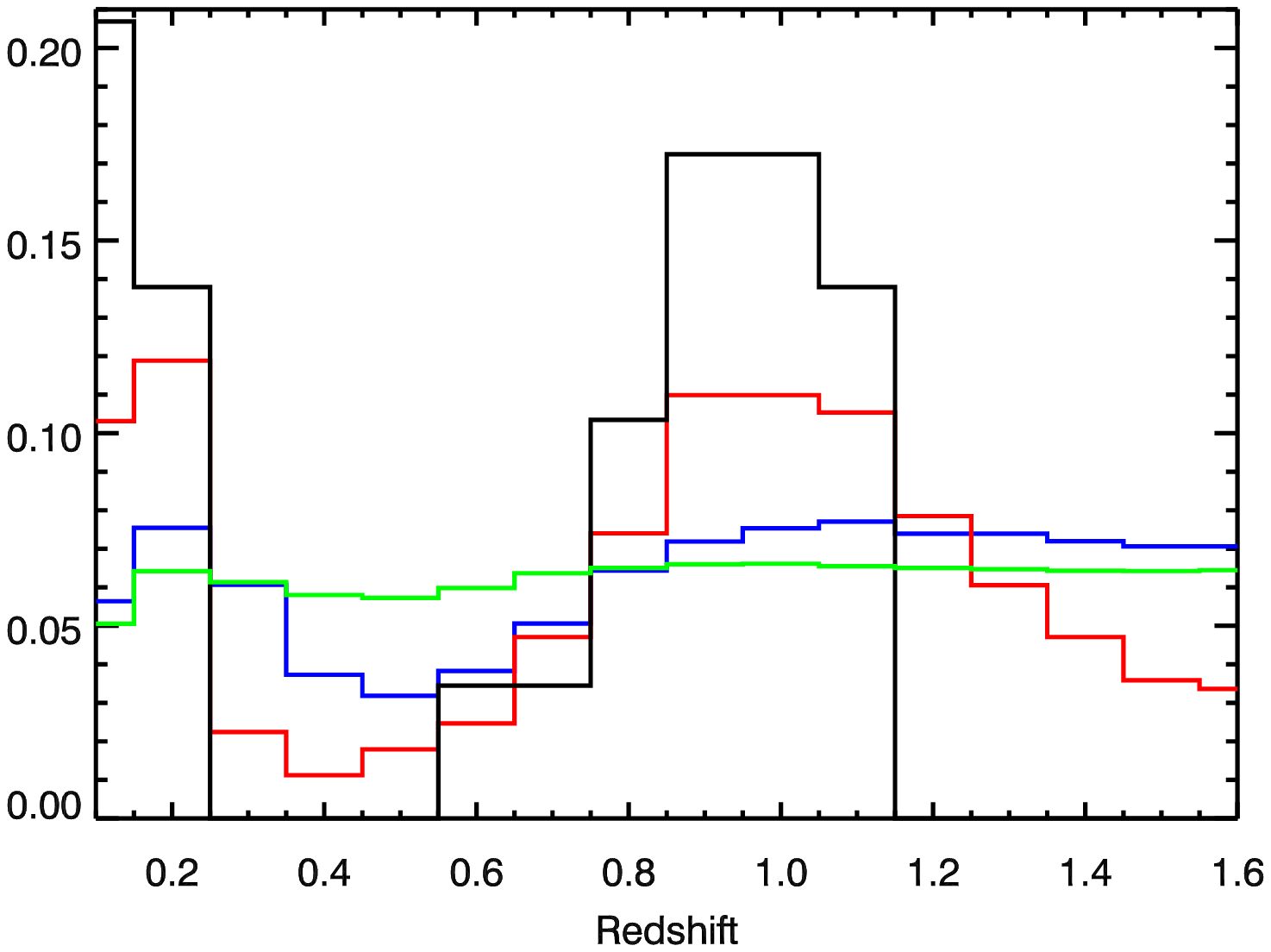} 
\includegraphics[width=\columnwidth]{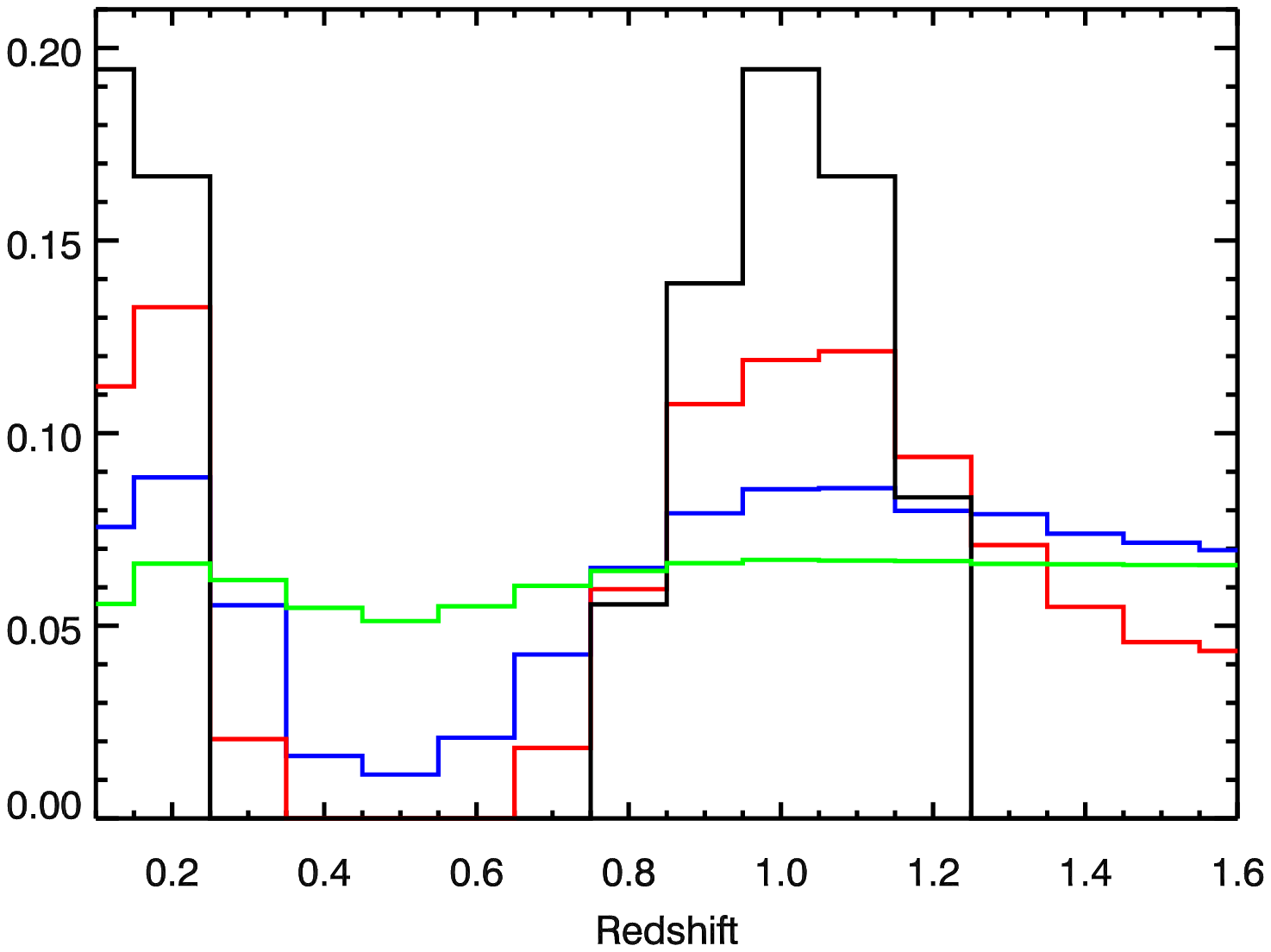} 
\end{centering} 
\caption{The histograms show the redshift locations of the most sensitive 
steps in the population function $f$ for various cuts in $\dcc$.  In 
order from highest to lowest peak the cuts are 90\%, 75\%, 50\%, 25\% 
of the maximum possible $\dcc$.  The top panel considers the $\dcc$ bias 
for the cosmology likelihood contour in the $\om$-$w_0$-$w_a$ space while 
the bottom panel is for the $w_0$-$w_a$ space.  Note that the most 
sensitive redshifts are robust to lower levels of bias, down to 50\% of 
maximum, and to the parameter space considered.} 
\label{fig:histtrs} 
\end{figure}

The location in redshift of the features in the population function 
also are important.  As we saw in Table~\ref{tab:poptrs}, $z\approx0.1$ 
and $z\approx1$ were key regions for sensitivity to bias.  In 
Fig.~\ref{fig:histtrs} we plot the redshift locations giving not just 
maximum bias, but greater than a certain percentage of maximum bias 
(still keeping full steps, i.e.\ $f=0$ or 1).  
We see that down to 50\% of the maximum possible bias the culprits are 
still population evolution around these sensitive redshifts.  This 
suggests that surveys designed to recognize population subclasses through  
especially comprehensive measurements around these redshifts can remove 
the top half of possible cosmology bias, improving the systematics by 
a factor two.

If we consider every possible form of the population function, randomly 
scanning over the number of steps and locations of transitions, then most 
of these will have little effect on the cosmology.  The mean bias, or 
$\langle\dcc\rangle$ will therefore be small.  For example, the mean is only 
0.08 for the $w_0$-$w_a$ contour.  This is simply due to combinatorics: 
there are many more ways of having, say, 8 steps over 17 bins than 1 step -- 
some 24000 times more possibilities -- and multistep functions will have 
little impact on the cosmology.  Furthermore the random location of the 
steps will also dilute the mean bias.  However, random population 
evolution is not the issue; for survey design we 
have to consider the worst case scenario, i.e.\ which systematics can 
give the most egregious misestimation of the cosmology results, and 
how to control this.  The results indicate that 
experiments should be guided by requirements to recognize subtypes 
with magnitude differences down to $\sim0.01$ mag (Fig.~\ref{fig:sigM}), 
and with particularly comprehensive measurements around $z\approx0.1$ 
and $z\approx1$ (Fig.~\ref{fig:histtrs}).

\section{Systematics: Dust Correction and Calibration \label{sec:dust}} 

The analysis of the constrained state function in terms of population 
drift was particularly straightforward because of the linear relation 
between the function $f(z)$ and the observable $m(z)$.  To illustrate 
a more complicated application we consider the systematic uncertainty 
due to dust extinction correction in supernova distances.  This is 
currently one of the dominant systematics 
\cite{conley,kowalski,hicken,nobili,goobar09} 
and uses measurements in 
multiple wavelength bands, or filters, to correct for the dust effects.  
However, if the different filters have some uncertainty 
in their calibrations then this propagates through to the relative fluxes 
or colors and then to the dust correction \cite{kimmiquel}.  
We use a simple, two band 
version of this as an illustration of a nonlinear, constrained systematic. 

We take the systematic to arise from zeropoint calibration errors in 
each filter, and the constraint can arise from subsidiary measurements 
such as on standard stars or instrumental calibration (see, e.g., 
\cite{stubbs}) that limit the zeropoint offsets to lie within 
$\Delta Z\in[Z_-,Z_+]$.  Again, it is the relative zeropoint differences, 
or colors, that cause bias; uniform offsets do not affect cosmology. 
Because the flux from sources at different redshifts peaks in different 
wavelength bands, the zeropoint errors will induce a redshift dependent 
error in the magnitude and hence a bias in the cosmology parameter 
estimation.  In addition, because the use of multiple bands to define 
the dust correction leads to an interdependence of SN at different 
redshifts, a correlated error matrix enters \cite{kimmiquel}. 

In the two band toy model for dust correction, the corrected magnitude 
$\tilde m$ is related to the magnitudes measured in two neighboring bands by 
\beq 
\tilde m_i=(1+R)\,m_i-R\,m_{i+1}\,, 
\eeq 
where $R$ is the extinction ratio.  We use as the two bands the 
restframe $B$ and $V$ bands for each supernovae, take $R=2.1$ 
(somewhat emphasizing the effect), 
and consider only calibration zeropoint error contributions to the 
dust correction, not any intrinsic SN color variation.  This simple 
model is sufficient to illustrate the effects of a nondiagonal error 
covariance matrix 
\beqa  
C&=&BEB^T \\ 
B&=&(1+R)\,\delta_{ij}-R\,\delta_{i,j-1}\,, 
\eeqa 
where $E$ is the pre-correction, possibly diagonal, error covariance 
matrix. 

The Fisher matrix is formed using the nondiagonal error covariance 
matrix $C$ and we then calculate the parameter bias due to zeropoint 
offsets $\Delta Z_k$ in the magnitudes by means of Eq.~(\ref{eq:biasgen}).  
We consider 8 filters logarithmically spaced in wavelength, with 
centers at $\lambda_0\,(1+a_\star)^{k-1}$, for $k=1-8$, taking 
$\lambda_0=4400$\AA\ and $a_\star=0.15$ so the maximum redshift 
corresponds to 1.66 \cite{omnibus}. 
Our state function is $\Delta Z$, which can be both positive and 
negative within the bounds, and we scan over all possible forms within 
the bounds and analyze the cosmology bias. 

Table~\ref{tab:bandtab} presents the results in the same format as the 
previous case in Table~\ref{tab:poptrs}.  However here the steps are 
in band zeropoints not population fractions and the locations are listed 
in terms of the filter numbers.  The important quantity is 
$\Delta Z_{\rm rel}$ between filters (recall that an overall zeropoint 
error has no cosmology effect), and needs to be constrained to the 
$\sim0.01$ level.  The furthest red filters, used for the highest 
redshift SN, are among the most sensitive to bias and should be tightly 
calibrated with instrumental and standard star measurements.

\begin{table}[h]
\begin{tabular}{cccc}
\hline 
Parameters \quad & Transitions \quad & max $\Delta\chi^{2}_{\Delta Z=0.01}$ 
\quad & $\Delta Z_{\rm rel}\,(1\sigma)$ \\ 
\hline 
$\Omega_{m}$ & 2-3, 7-8 & $4.07$ & 0.0098 \\ 
$w_{0}$ & 1-2, 4-5, 7-8 & $2.79$ & 0.012 \\ 
$w_{a}$ & 1-2, 4-5, 7-8 & $2.99$ & 0.012 \\ 
$w_{0},w_{a}$ & 3-4, 7-8 & $4.12$ & 0.015 \\ 
$\Omega_{m},w_{0},w_{a}$ & 3-4, 7-8 & $4.55$ & 0.018 \\ 
\hline 
\end{tabular}
\caption{For each set of parameters we consider the form of filter 
zeropoint errors that maximize the bias in terms of $\dcc$.  The 
transitions column gives the filter transitions of the maximizing 
function in $\Delta Z$, delivering a maximum bias $\dcc$ scaled to 
the case where the zeropoint calibration is bounded by $|\Delta Z|=0.01$, 
shown in the next column.  The last column shows the value of 
$\Delta Z_{\rm rel}$ {\it between\/} two filters that will shift the 
derived cosmology by $1\sigma$ from the true cosmology.} 
\label{tab:bandtab} 
\end{table}

As in the previous case in Sec.~\ref{sec:pop}, the shape, i.e.\ wavelength 
dependence of the zeropoint calibration, is unaffected by amplitude of the 
bounds on the calibration state function, but the 
absolute level is determined by the bounds.  If we consider twice as large
values for $\Delta Z$, then $\Delta\chi^2$ just scales with $\Delta Z^2$.

\begin{figure}[h]
\begin{centering}
\includegraphics[width=\columnwidth]{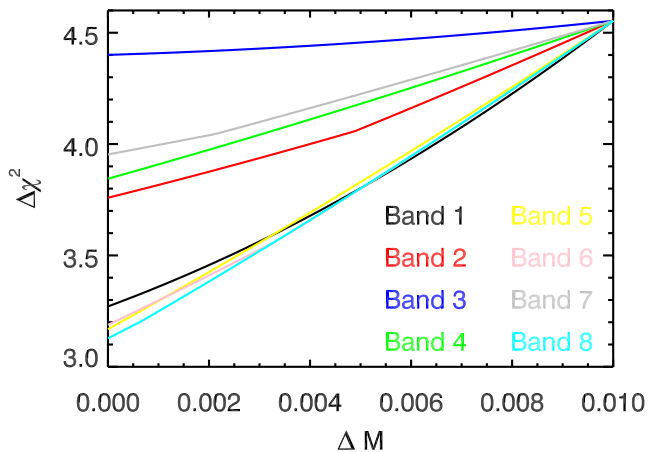} 
\end{centering} 
\caption{The maximum $\dcc$ bias is plotted as a function of the bound 
on individual filter band zeropoint offsets.  The greatest improvement in 
systematics control comes from improving the calibration uncertainty 
$\Delta Z$ for bands 1, 5, 6, 8 which correspond to the most cosmologically 
sensitive redshift leverages, near $z\approx0$, $1$, $1.7$.} 
\label{fig:eachband} 
\end{figure}

Any filter zeropoint step affects 
SN in some redshift range, and hence the overall cosmology.  Thus an 
improvement in the knowledge of a filter offset $\Delta Z_k$ reduces 
the maximum $\dcc$ bias.  Figure~\ref{fig:eachband} shows the effect 
of more tightly calibrating a given filter\footnote{We 
have also carried out PCA on the filter model Fisher matrix but find 
little useful information from the modes.  The wavelength band bin basis 
is better suited to actual design than saying, e.g., calibrate the 
combination of 0.4 times the first filter and $-0.2$ times the second 
filter etc.}.  
We see the greatest 
improvement for the end bands -- which are used for the lowest redshift, 
anchoring SN and the highest redshift, lever-arm SN -- and the middle 
bands, which are used near the sensitive $z\approx1$ region.  Thus survey 
design that provides particularly comprehensive calibration for these 
bands will see a large payoff in systematics control.

\section{Conclusions \label{sec:concl}}

Modeling unknown functions in cosmology is pervasive, whether 
these are functions that carry the physics we are directly interested 
in, such as the dark energy equation of state history, or functions 
describing subsidiary effects that we wish to subtract out, such as 
intermediate astrophysics or modeling of uncertainties.  If the 
functional form is assumed, then this becomes parameter fitting or 
``self-calibration'' but it is interesting and important to investigate 
the results when any viable function is allowed.  The viability is subject 
only to some constraints placed on the bounds of the function through 
theoretical or measurement input. 

We investigated two main issues for cosmological analysis in the presence 
of unknown but constrained functions.  First, we demonstrated a 
computationally efficient, pure and complete method for determining 
the viable space of principal component coefficients, with a simple 
geometric picture in terms of a hyperrectangle in the $N$-dimensional 
basis space.  This improves 
on the efficiency of the previous mode-by-mode method by a factor 3, 
while guaranteeing purity, i.e.\ validity of the selected functions. 
Conversely, compared to the also pure and complete direct scanning method, 
the projected hyperrectangle method gains in efficiency by factors of 
$10^{10}$ or greater. 

For many astrophysical problems the orthogonal bin basis (in redshift 
or wavelength) is well suited.  We evaluated two ``real world'' 
systematics issues with this method, one in redshift and one in wavelength. 
The first dealt with unrecognized population evolution in a subclass 
of standard candles.  Rather than assuming a form of evolution we 
analyzed the effects of all possible functional forms lying within some 
bounds.  The results provide quantitative guidance to controlling the 
worst cosmology biases arising from the systematic uncertainties.  In 
particular, we find that the regions $z\approx0.1$ and $z\approx1.0$ 
can benefit most from comprehensive observations to limit unrecognized 
subclasses (at the $\sim0.01$ mag level); surveys homogeneous over these 
redshift regions have improved control over cosmology misestimation. 

The second application concerned dust extinction corrections for Type Ia 
supernovae, where measurements in multiple wavelength bands can fit for 
dust but also correlate supernovae at different redshifts.  We analyzed 
the case where systematic uncertainties existed in the filter calibrations, 
bounded by instrumental or standard star observations, and propagated 
all possible functional forms into cosmology biases.  Again we found 
which specific forms were most damaging and that measurements designed 
to control such errors could remove the worst biases.  In particular, 
those bands used for the lowest and highest supernovae, and ones 
relevant around $z\approx1.0$, should be most comprehensively calibrated 
(to the $\sim0.01$ mag level relatively) for a more robust survey.

\acknowledgments 

We are grateful for useful discussions with Marina Cort\^{e}s, Alex Kim, 
Saul Perlmutter, and especially Roland de Putter.  
JS acknowledges support from the OTICON Fund and Dark Cosmology Centre, 
and thanks the Berkeley Center for Cosmological Physics and Berkeley 
Lab for hospitality during his stay. 
This work has been supported in part by the Director, Office of Science, 
Office of High Energy Physics, of the U.S.\ Department of Energy under 
Contract No.\ DE-AC02-05CH11231.

\section*{Appendix: Efficient Projection and Boundary 
Definition \label{sec:apx}} 

Since the dark energy physics of interest mostly translates into 
descriptions of the EOS in redshift bin space, our bounds on the 
state function are given in this space.  In this Appendix we will 
detail, through a geometrical understanding, a fast method for 
analyzing how these bounds in bin space may be carried into bounds 
on coefficients in principal components space.  The bounds on $w(z)$ 
directly form the space of the permitted sets of these coefficients, 
but picking a pure and complete set of coefficients in PC space is 
not so simple. 

The value for the EOS, say, at a particular redshift $z$ is not 
described by just one mode as in the bin space case, but by a linear 
combination of many modes weighted by their respective coefficients.  
To scan the whole coefficient space within the bounds given in 
redshift space requires an impractical amount of computational power, 
as discussed in Sec.~\ref{sec:method}.  We need a more clever method 
of obtaining the desired results. 

For a set of independent state function bounds on $\{w_{i}\}$, 
the permitted space in bin space is bounded by a hyperrectangle, or 
orthotope.  The coordinates of the corners are given by any combination 
$W_{1}^{\pm},W_{2}^{\pm}, \dots ,W_{N}^{\pm}$ where $W_{i}^{\pm}$ 
denotes the maximum and minimum values allowed for bin $i$, and $N$ 
is the number of bins. This fixes the $2^{N}$ corners.  This orthotope 
structure contains all the needed information on the boundary of the 
allowed space no matter the basis.  

Let us first address how to determine the viable region for any set 
of principal component coefficients, $\alpha_{1},\dots,\alpha_{m}$.  
The permitted space is simply the projection of the orthotope onto the
subspace spanned by these components.  The nodes of the boundary in 
PC space are found by projecting the vectors going from the origin to 
the corners of the orthotope.  Denoting these ($N$-dimensional) vectors 
as $\mathbf{c_{i}}$ and the projected ($m$-dimensional) node vectors 
in the $\mathbf{S}=[e_{1},\dots,e_{m}]$ PC space as 
$\mathbf{p_{i}}=(p_{i1},\dots,p_{im})$, where $p_{i1},\dots,p_{im}$ 
are the coordinates of $\mathbf{p_{i}}$ with respect to the axes 
$\alpha_{1},\dots,\alpha_{m}$, we find 
\begin{equation}
\mathbf{p_{i}}=(\mathbf{S}^{T}\mathbf{S})^{-1}\mathbf{S}^{T}\mathbf{c_{i}} \,.\label{eq:2}
\end{equation} 

Consider now the case where we keep only a subset of PC modes, i.e.\ we 
do not marginalize over the other modes but fix their coefficients, 
e.g.\ to 0.  
Information is lost by not using the complete basis in the expansion, 
but sometimes data or practicalities do not enable us to know or measure 
all modes. Therefore, we must sometimes sample with only a subset $M$ of 
modes.  The new permitted space of PC coefficients is the 
intersection of the $M$-dimensional space with the original 
$N$-dimensional space.  This subspace is completely defined by the 
boundary points, found by looking at the intersections
between the $M$-dimensional space of interest and the $N$-dimensional 
bounded bin space (cf.\ the black dots in Fig.~\ref{fig:subspace}).  
However, the allowed space is not generally a 
$M$-rectangle; for example, a plane cutting through a cube can have 
a boundary with six corners unless it is specially oriented.  This has 
the consequence that there is no $M$-dimensional basis that will generally 
be pure and complete (i.e.\ there are no orthogonal axes spanning the 
space).  Only the 
original bin basis in $N$-dimensions can provide a pure and complete 
description of the valid state functions. 

To describe the actual procedure for evaluating the allowed region in 
PC space we rephrase the issue more mathematically.  The determination 
of the coefficient bounds is 
solved by considering the intersection of the subspace $\mathbf{C_{a}}$,
spanned by the principal components keeping the selected number of
coefficients constant, with the orthotope.  
In the $3$ dimensional case, with a 2 dimensional subspace, 
the boundaries are of $3-2$ dimensions, i.e.\ lines, having parametric form 
$\mathbf{I}_{a}+(\mathbf{I}_{b}-\mathbf{I}_{a})t$ and the plane is 
parameterized by $\mathbf{P}_{0}+(\mathbf{P}_{1}-\mathbf{P}_{0})u+ 
(\mathbf{P}_{2}-\mathbf{P}_{0})v$, where $t$, $u$, $v$ are real numbers, 
and $\mathbf{I}_{i}$ and $\mathbf{P}_{j}$ are points on the line and in 
the plane, respectively.  Setting 
$\mathbf{D}=[\mathbf{I}_{a}-\mathbf{P}_{0}]$,
$\mathbf{A}=[\mathbf{I}_{a}-\mathbf{I}_{b},\mathbf{P}_{1}-\mathbf{P}_{0},\mathbf{P}_{2}-\mathbf{P}_{0}]$, 
and $\mathbf{L}=[t,u,v]$ we find 
\begin{equation} 
\mathbf{D^{-1}A=L}\,. 
\end{equation} 
In the higher dimensional case the solution has the identical form.  
Geometrically, $\mathbf{I}$ represents vectors going from the origin 
to the corners of the orthotope (so $t\in[0,1]$) and $\mathbf{P}$ are 
\emph{non-}colinear points in $\mathbf{C_{a}}$.  The matrix 
manipulations can be computed easily so solving for the allowed 
region in PC space is highly efficient and quick.

\end{document}